 \documentclass{aastex}

\newcommand{\hsp}{\hspace*}

\newcommand{\E}[1]{\mbox{$\times10^{#1}$}}

\newcommand{\lumi}{\mbox{erg s$^{-1}$}}
\newcommand{\flux}{\mbox{erg s$^{-1}$ cm$^{-2}$}}

\newcommand{\srci}{\mbox{CXOM82~J095554.7+694101}}
\newcommand{\srcii}{\mbox{CXOM82~J095552.3+694054}}
\newcommand{\srciii}{\mbox{CXOM82~J095551.6+694036}}
\newcommand{\srciv}{\mbox{CXOM82~J095551.4+694044}}
\newcommand{\srcv}{\mbox{CXOM82~J095551.1+694045}}
\newcommand{\srcvi}{\mbox{CXOM82~J095550.7+694044}}
\newcommand{\srcvii}{\mbox{CXOM82~J095550.2+694047}}
\newcommand{\srcviii}{\mbox{CXOM82~J095547.5+694036}}
\newcommand{\srcix}{\mbox{CXOM82~J095546.6+694041}}

\slugcomment{Accepted by ApJL}

\shorttitle{Luminous Variable Source in M82}
\shortauthors{Matsumoto et al.}

\begin{document}


\title{Discovery of a Luminous, Variable, Off-Center Source 
in the Nucleus of M82 with the {\it Chandra} HRC}


\author{H. Matsumoto\altaffilmark{1}, 
T. G. Tsuru\altaffilmark{2}, 
K. Koyama\altaffilmark{2,3}, 
H. Awaki\altaffilmark{4},
C. R. Canizares\altaffilmark{1}, 
N. Kawai\altaffilmark{5},\\
S. Matsushita\altaffilmark{6}, 
A. Prestwich\altaffilmark{7}, 
M. Ward\altaffilmark{8}, 
A. L. Zezas\altaffilmark{7},
and
R. Kawabe\altaffilmark{9}
}


\altaffiltext{1}{Center for Space Research, Massachusetts Institute of Technology, 77 Massachusetts Avenue, Cambridge, MA 02139-4307, USA; e-mail(HM) matumoto@space.mit.edu}
\altaffiltext{2}{Department of Physics, Faculty of Science, Kyoto University, Sakyo-ku, Kyoto 606-8502, Japan}
\altaffiltext{3}{CREST: Japan Science and Technology Corporation (JST), 4-1-8 Honmachi, Kawaguchi, Saitama 332-0012, Japan}
\altaffiltext{4}{Faculty of Science, Ehime University, 2-5 Bunkyo-cho, Matsuyama 790-8577, Japan}
\altaffiltext{5}{RIKEN (The Institute of Physical and Chemical Research),
2-1 Hirosawa, Wako, Saitama 351-0198, Japan}
\altaffiltext{6}{Submillimeter Array, Harvard-Smithsonian Center for 
Astrophysics, 60 Keawe St. Suite 204, Hilo, HI 96720, USA}
\altaffiltext{7}{Harvard-Smithsonian Center for Astrophysics, 60 Garden Street, Cambridge, MA 02138, USA}
\altaffiltext{8}{X-ray Astronomy Group, Department of Physics and Astronomy, University of Leicester, Leicester LE1 7RH, UK}
\altaffiltext{9}{Nobeyama Radio Observatory, Minamimaki, Minamisaku, Nagano 384-1305, Japan}


\begin{abstract}
We present results from observations of the most famous
starburst galaxy M82 with the High-Resolution Camera onboard
the {\it Chandra X-Ray Observatory}.  We found nine sources
in the central $1\arcmin\times1\arcmin$ region, but no
source was detected at the galactic center.  Comparing the
observations on 1999 October 28 and on 2000 January 20, 
we found four of the nine sources showed significant time
variability.  In particular, {\srcvii}, which is $9\arcsec$
away from the galactic center, showed extremely large time
variability. We conclude that this source is the origin of
the hard X-ray time variability of M82 detected with {\it
ASCA}.  Assuming a spectral shape obtained by the {\it ASCA}
observation, its luminosity in the 0.5 -- 10 keV band
changed from 1.2\E{40} {\lumi} on 1999 October 28 to
8.7\E{40} {\lumi} on 2000 January 20.

\end{abstract}




\section{Introduction}

The X-ray spectrum of the starburst galaxy M82 measured with
{\it ASCA} {\citep{Tanaka1994}} consists of soft, medium,
and hard components {\citep{Tsuru1997}}.  The hard
component, which dominates the X-ray spectrum above 2 keV,
was found to show time variability by monitoring M82 with
{\it ASCA} in 1996 {\citep{Matsumoto1999,Ptak1999}}.
Assuming a distance of 3.9 Mpc (i.e. $1\arcsec \sim$ 19 pc)
{\citep{Sakai1999}}, its luminosity in the 0.5 -- 10 keV
band changed between 4.5\E{40} {\lumi} and 1.6\E{41} {\lumi}
at various time scales from 1\E{4}~s to a month.  The most
plausible explanation for the variability is that a
low-luminosity AGN exists in M82. If this explanation is
correct, M82 would be one of the most important objects for
the study of a relation between an AGN and starburst
activity {\citep[e.g.][]{Umemura1997}} because of its
proximity.  However, hard X-ray spectra of M82 obtained with
{\it Ginga} and {\it BeppoSAX}, which can detect the X-ray
photons of higher energy than {\it ASCA}, differ from
typical AGN spectra, since they cannot be fitted by a
power-law model {\citep{Tsuru1992,Cappi1999}}.  Therefore,
it is also possible that the origin of the hard component is
a new type of compact X-ray source.

Furthermore, {\citet{Matsumoto1999}} compared the {\it ASCA}
image with the {\it ROSAT} HRI image
{\citep{Strickland1997}}, and showed that the hard component
probably comes from the X-ray brightest source detected with
the {\it ROSAT} HRI.  The HRI source was found to show time
variability {\citep{Collura1994}}. Since
{\citet{Stevens1999}} showed that the source is away from
the dynamical center of M82, this galaxy may harbor an
``off-center'' AGN.

Thus, the source of the hard component of M82 would be a
quite interesting and important object. For further
investigation, it is necessary to determine the position of
the hard component precisely and to find a counterpart in
other wavelengths if possible.  Therefore, we analyzed the
data of M82 obtained with the High-Resolution Camera (HRC)
{\citep[e.g.][]{Murray1987,Murray1997}} onboard the {\it
Chandra X-Ray Observatory (CXO)}{\citep{Weisskopf1995}}. The
HRC has sensitivity in the 0.1 -- 10 keV band with a peak at
about 1 keV, and it has the highest angular resolution (FWHM
$\sim0\farcs5$) of all instruments onboard X-ray
observatories to date. Uncertainties in this paper refer to
1$\sigma$ confidence limits.

\section{Data Analysis and Results}

M82 was observed twice with {\it CXO} as a calibration
target using the HRC-I on 1999 October 28 and 2000 January
20.  The HRC data were processed with the standard procedure
by the {\it Chandra X-ray Center (CXC)}
{\footnote{\url{http://asc.harvard.edu/}}}, and we used only
the fully processed science products (level = 2) as event
files.  We also applied an event screening procedure
developed by the HRC team to eliminate ``ghost'' events.
The exposure times are 2788~s for the first observation and
17684~s for the second.

\begin{figure}
\figurenum{1}
(a)\hsp{7cm}(b)\\
\plottwo{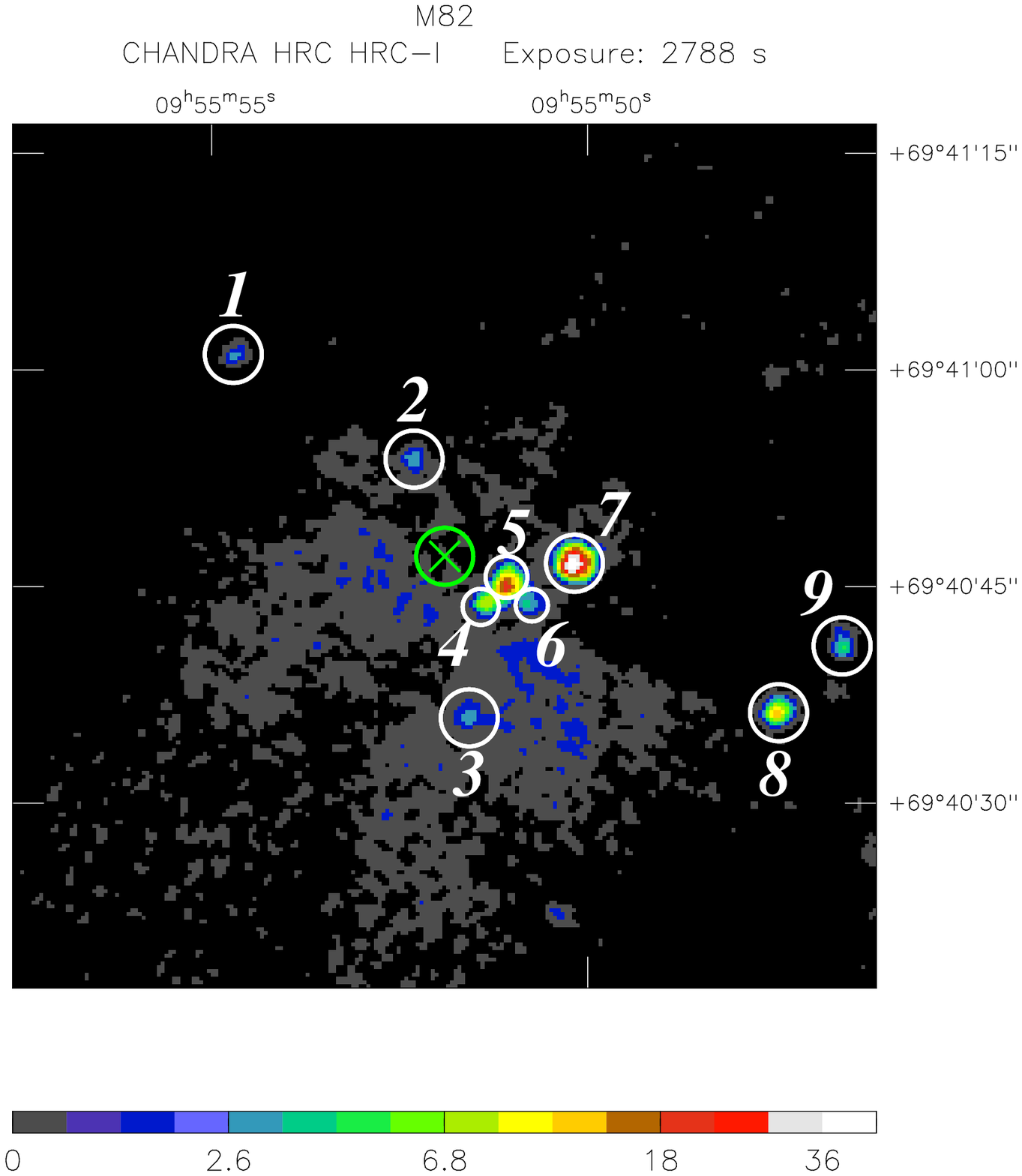}{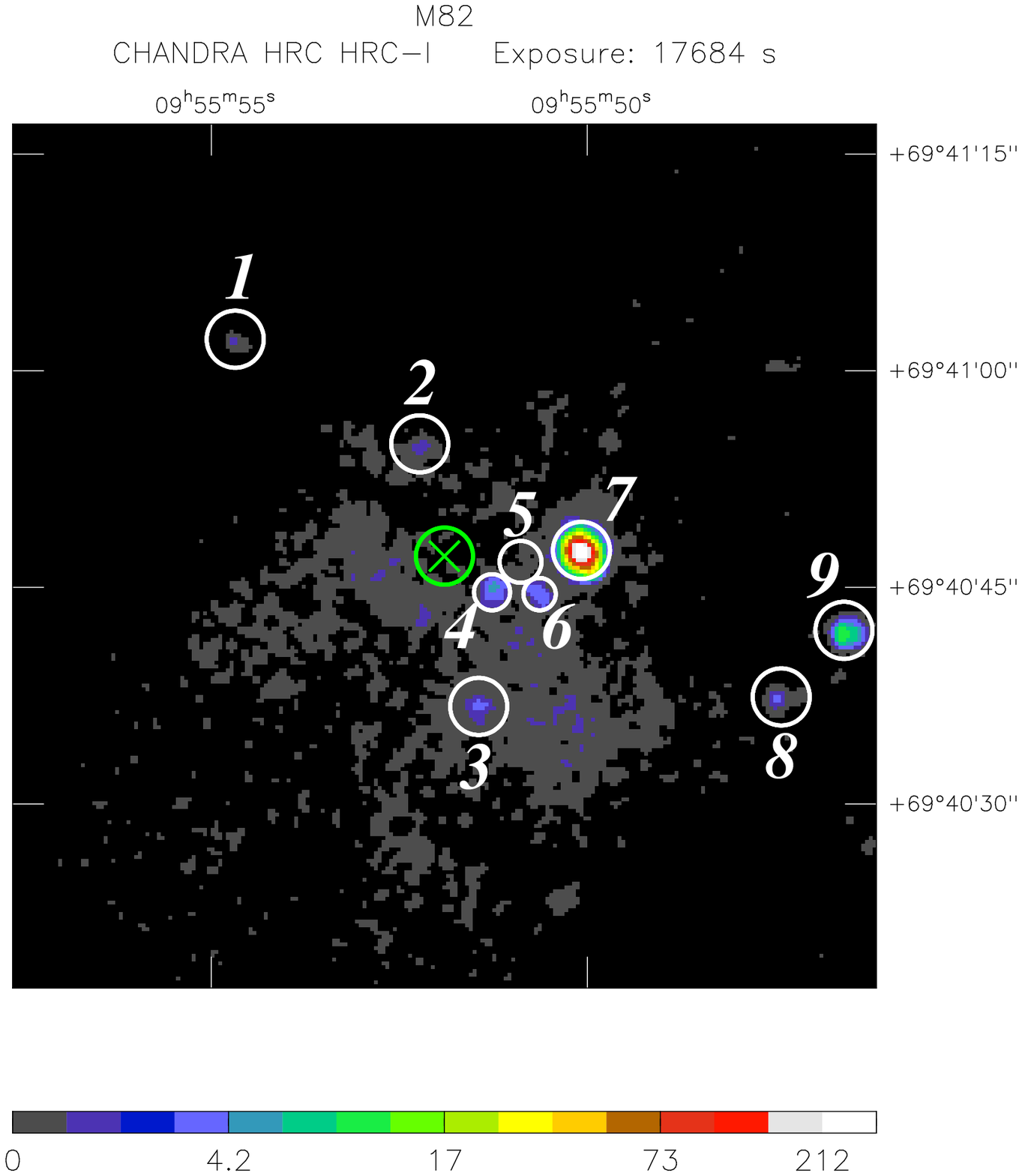}
\caption{
Central $1\arcmin\times1\arcmin$ (= 1.1
kpc $\times$ 1.1 kpc) region of M82: (a) 1999 October 28,
(b) 2000 January 20.  The pixel size is $0\farcs26$ and the
image has been smoothed with a Gaussian filter with a
$\sigma$ of 1 pixel. The color scale shows the total count
per pixel.  The green cross shows the radio kinematic
center, and the green circle is its position error circle
with a radius of $2''$ {\citep{Weliachew1984}}.  The sources
we analyzed in this paper are designated with white circles
and numbers. The circular regions were used to extract X-ray
events for the sources.
\label{fig:HRC}
}
\end{figure}

We found that all X-ray bright sources which can be the
origin of the {\it ASCA} hard component exist within the
central $1\arcmin\times1\arcmin$ (= 1.1 kpc $\times$ 1.1
kpc) region, which is shown in Figure~{\ref{fig:HRC}}.  The
position of the bright off-center source found with the {\it
ROSAT} HRI {\citep{Stevens1999}} is also within the field.
Therefore, we concentrate on the central
$1\arcmin\times1\arcmin$ region.  In Figure~\ref{fig:HRC},
we see nine prominent sources, which are designated with
circles and numbers.  More detailed analysis including
diffuse emission and other fainter sources will be presented
in a forthcoming paper (M. Ward et al., in preparation).  We
determined the positions of the peaks of the nine HRC
sources with the wavelet algorithm using {\it Chandra
Interactive Analysis of Observations (CIAO)}, and named the
HRC sources by using these positions (Table~{\ref{tbl:src}}).
The position uncertainty is less than $\sim 0\farcs1$.

According to ``the {\it CXC} memo on astrometry problems''
\footnote{\url{http://asc.harvard.edu/ciao/caveats/aspect4.html}},
the HRC event files we used could have offsets up to
$10\arcsec$ in the celestial coordinates.  To check the
reliability of the coordinates in the event files, we
compared the coordinates of the HRC sources with the Two
Micron All Sky Survey (2MASS) Point Source Catalog. There
are 17 2MASS sources in the field of
Figure~{\ref{fig:HRC}}. According to the {\it CXC}, the
offset of the celestial coordinate in ordinary CXO event
files is $1\arcsec$ (RMS).  The position uncertainty in the
2MASS Catalog is $\sim 0\farcs1$.  Considering these
uncertainties, we found that three 2MASS sources out of 17
agree with the HRC sources (Nos.~1, 6, and 8).  Since the
probability of a chance coincidence is less than 0.1~\%, we
can assume that the HRC coordinates are reliable. We then
compared the coordinates of the HRC sources with the 5~GHz
sources {\citep{Muxlow1994}} whose position accuracy is
$\sim0\farcs5$. The fact that four HRC sources (Nos.~4, 5,
6, and 7) have counterparts on the 5~GHz map also supports
the reliability of the HRC coordinates. The identifications
of the HRC sources are shown in Table~{\ref{tbl:src}}.

We estimated the counting rates of the HRC sources using the
X-ray events within the circular regions shown in
Figure~\ref{fig:HRC}.  The backgrounds including the diffuse
emission were estimated using source-free regions around the
sources.  The counting rates are listed in
Table~\ref{tbl:src}. Four HRC sources (Nos.~5, 7, 8, and 9)
show significant time variability. In particular, No.~5
disappeared in the second observation, and No.~7 became
brighter by a factor of 7.  The counting rates of No.~1 and
No.~4 did not change significantly between the two
observations.

The dynamical center at $(\alpha, \delta)_{\rm J2000}$ =
$(9^{\rm h}55^{\rm m}51.9^{\rm s},
69{\degr}40{\arcmin}47.1{\arcsec})$, which is determined by
the radio observation of the motion of \ion{H}{1} gas, is
shown as the green cross in Figure~\ref{fig:HRC}
{\citep{Weliachew1984}}.  The position error circle with a
radius of $2\arcsec$ is also shown as the green circle.  We
should note that all nine HRC sources are clearly away from
the dynamical center.  The counting rate of the dynamical
center estimated with the extraction radius of $2\arcsec$ is
($1.79\pm1.48$)\E{-3} c s$^{-1}$ for 1999 October 28, and
($1.12\pm6.84$)\E{-4} c s$^{-1}$ for 2000 January 20. Thus,
we found no significant source at the dynamical center.

We estimated the HRC counting rate of the {\it ASCA} hard
component {\citep{Matsumoto1999}} using {\it W3PIMMS} v3.0
\footnote{\url{http://heasarc.gsfc.nasa.gov/Tools/w3pimms.html}}: the
expected counting rate is 0.72 c s$^{-1}$ in the highest
state and 9.7\E{-2} c s$^{-1}$ in the lowest. Only the
counting rate of No.~7 is consistent with the expected
counting rate. Furthermore, the position of the {\it ASCA}
hard component is consistent with that of No.~7
{\citep{Matsumoto1999}}. Therefore, we can conclude that the
variability of the {\it ASCA} hard component is due to No.~7
({\srciv}).  The separation between No.~7 and the dynamical
center is $9\arcsec$ ($\sim$ 170 pc).

\section{Discussion}

We found that 41.5+59.7 {\citep{Kronberg1985}} is a
candidate of the radio counterpart of No.~7 as well as
41.30+59.6 {\citep{Muxlow1994}}.  41.5+59.7 is $0\farcs76$
away from No.~7, while the separation between 41.30+59.6 and
No.~7 is $0\farcs96$. According to the morphology and
spectral shape in the 5~GHz radio band, {\citet{Muxlow1994}}
suggested that 41.30+59.6 is a young supernova remnant
(SNR).  41.5+59.7 show a 100~\% drop in the radio flux
within a year, and its radio decay time scale and spectrum
are very similar to SN~1983n
{\citep{Kronberg1985b,Kronberg2000}}. Therefore, this source
may also be a SNR.  The hard X-ray emission of M82 was found
to show the short-term variability
{\citep{Matsumoto1999,Ptak1999}}.  Since it is rather
difficult to explain the short-term variability in terms of
a SNR origin, both radio sources may not be a real
counterpart of No.~7.

If we assume the spectral shape of No.~7 is an absorbed
thermal bremsstrahlung model with a temperature of 10 keV
and a column density of $10^{22}$ cm$^{-2}$, which is the
typical spectral shape of the {\it ASCA} hard component
{\citep{Matsumoto1999}}, the peak X-ray flux of No.~7 is
6.6\E{-13} {\flux} in the 2 -- 10 keV band. According to the
the $\log N - \log S$ relation {\citep{Ueda1999}}, the
probability that a source as bright as No.~7 exists in the
$1\arcmin\times1\arcmin$ field is $\sim$ 0.3 \%.  Therefore,
No.~7 is probably not a background AGN.  The {\it ASCA}
spectrum of the variable source obtained by subtracting the
spectrum of the lowest state from the highest state shows
heavy absorption (the column density is $\sim$
$10^{22}$~cm$^{-2}$) {\citep{Matsumoto1999}}. Since the
Galactic absorption toward M82 is 4\E{20}~cm$^{-2}$
{\citep{Dickey1990}}, the variable source is embedded deeply
in M82, and hence No.~7 is probably not a foreground source,
unless the source has extremely large intrinsic absorption.

Assuming the absorbed bremsstrahlung model of 10 keV and
$10^{22}$ cm$^{-2}$, unabsorbed X-ray luminosity in the 0.5
-- 10 keV band ($L_{\rm X}^{\rm 0.5-10 keV}$) is expressed
using an HRC counting rate ($C_{\rm HRC}$) as
\begin{equation}
L_{\rm X}^{\rm 0.5-10 keV}
 = 1.67\E{38}\ \mbox{erg s$^{-1}$} \times 
\left[\frac{C_{\rm HRC}}{10^{-3}\ \mbox{c s$^{-1}$}}\right].
\end{equation}
The $L_{\rm X}^{\rm 0.5-10 keV}$ of No.~7 was estimated to
be 1.2\E{40} {\lumi} on 1999 October 28 and 8.7\E{40}
{\lumi} on 2000 January 20.

If we assume that No.~7 is a black hole (BH) and that the
maximum luminosity does not exceed the Eddington luminosity,
the mass of the BH must be larger than 700 $M_\sun$, and
No.~7 is not a stellar-mass BH ($\sim$ 10 $M_\sun$).  Since
No.~7 is $9\arcsec$ away from the dynamical center, the mass
of No.~7 must be much smaller than the gravitational mass
within $9\arcsec$ from the center which is 4\E{8} $M_\sun$
{\citep{McLeod1993}}, otherwise the dynamical center would
be shifted from the current position. Therefore, No.~7 is at
the low end of the mass distribution of super-massive BHs
($10^6$ -- $10^9$ $M_\sun$) or a medium-massive BH ($10^3$
-- $10^6$ $M_\sun$).  The possibility of the medium-massive
BH is discussed in detail in {\citet{Matsushita2000b}} along
with the discovery of an expanding molecular superbubble
surrounding No.~7 {\citep{Matsushita2000}}. Other
possibilities such as an X-ray binary source whose jet is
strongly beamed at us cannot be excluded. Further
investigation including other wavelengths is strongly
encouraged to reveal the true character of No.~7 (\srcvii).

If we use equation (1) to the sources other than No.~7,
their $L_{\rm X}^{\rm 0.5-10 keV}$s are much greater than
the Eddington luminosity for a 1.4 $M_\sun$
object. Therefore, these sources may also be BHs.

Though we found no significant source at the dynamical
center, it is still possible that a faint X-ray source such
as Sgr A$^*$ exists {\citep{Koyama1996}}. Assuming an
absorbed power-low model with a photon index of 1.7 and a
column density of $10^{22}$ cm$^{-2}$, the upper limit of
$L_{\rm X}^{\rm 0.5-10 keV}$ of the dynamical center is
5.4\E{38} {\lumi} for 1999 October 28 and 1.3\E{38} {\lumi}
for 2000 January 20.



\acknowledgments

This paper makes use of data products from the Two Micron
All Sky Survey, which is a joint project of the University
of Massachusetts and the Infrared Processing and Analysis
Center/California Institute of Technology, funded by the
National Aeronautics and Space Administration and the
National Science Foundation. The authors are grateful to
Miss Deborah Gage for careful review of the manuscript. We
also thank Dr. P. Kaaret for valuable comments.  H. M. and
S. M. are supported by the JSPS postdoctoral Fellowships for
Research Abroad.





\clearpage


\begin{deluxetable}{ccccccc}
\tabletypesize{\scriptsize}
\tablecaption{X-ray sources at the central region of M82 detected with the HRC. \label{tbl:src}}
\tablewidth{0pt}
\tablehead{
\colhead{Number}&\colhead{Source Name\tablenotemark{a}}
&\multicolumn{2}{c}{Identification}
&\multicolumn{3}{c}{Counting Rate ($10^{-3}$ c s$^{-1}$)}\\
\cline{3-4}\cline{5-7}\\
&&\colhead{Infrared\tablenotemark{b}}&\colhead{Radio\tablenotemark{c}}
&\colhead{$R$\tablenotemark{d}}&\colhead{1999 Oct 28}&\colhead{2000 Jan 20}
}
\startdata
1&\srci		&0955547+694100		&\nodata
&$2\farcs0$	&2.51$\pm$1.08	&2.42$\pm$0.40\\

2&\srcii	&\nodata		&\nodata
&$2\farcs0$	&6.10$\pm$2.00	&3.15$\pm$0.67\\

3&\srciii	&\nodata		&\nodata
&$2\farcs0$	&4.30$\pm$2.43	&8.43$\pm$0.95\\

4&\srciv	&\nodata		&42.65+57.8	
&$1\farcs3$	&8.04$\pm$2.14	&6.27$\pm$0.72\\

5&\srcv		&\nodata		&42.21+59.0	
&$1\farcs4$	&20.9$\pm$3.1	&0.292$\pm$0.505\\

6&\srcvi	&0955507+694043		&41.95+57.5	
&$1\farcs1$	&2.99$\pm$1.54	&5.20$\pm$0.65\\

7&\srcvii	&\nodata		
&41.30+59.6 or 41.5+59.7\tablenotemark{e}	
&$2\farcs0$	&71.1$\pm$5.9	&520$\pm$5\\

8&\srcviii	&0955475+694036		&\nodata	
&$2\farcs0$	&24.8$\pm$3.1	&4.89$\pm$0.66\\

9&\srcix	&\nodata		&\nodata	
&$2\farcs0$	&4.63$\pm$1.48	&22.8$\pm$1.2\\

\nodata &dynamical center\tablenotemark{f}		
&\nodata		&\nodata
&$2\farcs0$	&1.79$\pm$1.48	&0.112$\pm$0.684\\
 \enddata


\tablenotetext{a}
{Sources are named according to the Chandra Source Naming Convention
(\url{http://asc.harvard.edu/udocs/naming.html}).\\
For example, CXOM82~J095554.7+694101 is a source at 
($\alpha$, $\delta$)$_{\rm J2000}$=$(9^{\rm h}55^{\rm m}54.7^{\rm s},
69{\degr}41{\arcmin}01{\arcsec})$.}
\tablenotetext{b}{The Two Micron All Sky Survey (2MASS) Point Source Catalog
(\url{http://www.ipac.caltech.edu/2mass/})}
\tablenotetext{c}{\citet{Muxlow1994}}
\tablenotetext{d}{Extraction radius used to estimate the counting rate.
The radii for Nos.~4, 5, and 6 were limited to below
$2\arcsec$, because they are close to each other.  }
\tablenotetext{e}{\citet{Kronberg1985}}
\tablenotetext{f}{$(\alpha, \delta)_{\rm J2000}$ = 
$(9^{\rm h}55^{\rm m}51.9^{\rm s},
69{\degr}40{\arcmin}47.1{\arcsec})$
\citep{Weliachew1984}}

\tablecomments{Errors for the counting rates are given 
for 1$\sigma$ confidence.}

\end{deluxetable}

\clearpage



\begin{thebibliography}{}

\bibitem[Cappi et al.(1999)]{Cappi1999}
Cappi, M. et al., 1999, \aap, 350, 777

\bibitem[Collula et al.(1994)]{Collura1994}
Collura, A., Reale, F., Schulman, E., Bregman, J. N. 1994,
\apj, 420, L63

\bibitem[Dickey \& Lockman(1990)]{Dickey1990}
Dickey, J. M., \& Lockman, F. J. 1990, \araa, 28, 215

\bibitem[Koyama et al.(1996)]{Koyama1996}
Koyama, K., Maeda, Y., Sonobe, T., Takeshima, T., Tanaka,
Y., \& Yamauchi, S. 1996, \pasj, 48, 249

\bibitem[Kronberg, Biermann, \& Schwab(1985)]{Kronberg1985}
Kronberg, P. P., Biermann, P., \& Schwab, F. R. 1985, \apj, 291, 693

\bibitem[Kronberg \& Sramek(1985)]{Kronberg1985b}
Kronberg, P. P., \& Sramek, R. A. 1985, Science, 227, 28

\bibitem[Kronberg et al.(2000)]{Kronberg2000}
Kronberg, P. P., Sramek, R. A., Birk, G. T., Dufton, W.,
Clarke, T. E., Allen, M. L. 2000, \apj, 535, 706

\bibitem[McLeod et al.(1993)]{McLeod1993} 
McLeod, K. K., Rieke, G. H., Rieke, M. J., \& Kelly,
D. M. 1993, \apj, 412, 111

\bibitem[Matsumoto \& Tsuru(1999)]{Matsumoto1999}
Matsumoto, H., \& Tsuru, T. G. 1999, \pasj, 51, 321

\bibitem[Matsushita(2000)]{Matsushita2000} 
Matsushita, S. 2000, Ph.D. thesis, The Graduate University
for Advanced Studies, Japan

\bibitem[Matsushita et al.(2000)]{Matsushita2000b}
Matsushita, S. et al. 2000, \apjl, submitted

\bibitem[Murray et al.(1987)]{Murray1987}
Murray, S. S., Chappell, J. H., Elvis, M. S., Forman, W. R.,
\& Grindlay, J. E. 1987, Astrophys. Lett. Commun., 26, 113

\bibitem[Murray et al.(1997)]{Murray1997}
Murray, S. S., et al. 1997, Proc. SPIE, 3114, 11

\bibitem[Muxlow et al.(1994)]{Muxlow1994} 
Muxlow, T. W. B., Pedlar, A., Wilkinson, P. N., Axon, D. J., 
Sanders, E. M. \& de Btuyn, A. G. 1994, \mnras, 266, 455

\bibitem[Ptak \& Griffiths(1999)]{Ptak1999}
Ptak, A., \& Griffiths, R. 1999, \apj, 517, L85

\bibitem[Sakai \& Madore(1999)]{Sakai1999}
Sakai, S., \& Madore, B. F., 1999, \apj, 526, 599

\bibitem[Stevens, Strickland, \& Wills(1999)]{Stevens1999}
Stevens, I. R., Strickland, D. K., \& Wills, K. A., 1999, \mnras,
308, L23

\bibitem[Strickland, Ponman, \& Stevens(1997)]{Strickland1997}
Strickland, D. K., Ponman, T. J., \& Stevens, I. R.,
1997, \aap, 320, 378

\bibitem[Tanaka, Inoue, \& Holt(1994)]{Tanaka1994}
Tanaka, Y. Inoue, H., \& Holt, S. S. 1994, \pasj, 46, L37

\bibitem[Tsuru(1992)]{Tsuru1992}
Tsuru T. G. 1992, Ph.D. thesis, The University of Tokyo

\bibitem[Tsuru et al.(1997)]{Tsuru1997}
Tsuru T. G., Awaki H., Koyama K., Ptak A. 1997, \pasj, 49,
619

\bibitem[Ueda et al.(1999)]{Ueda1999}
Ueda, Y. et al. 1999, \apj, 518, 656

\bibitem[Umemura, Fukue, \& Mineshige(1997)]{Umemura1997}
Umemura, M., Fukue, J., \& Mineshige, S., 1997, \apj, 479, L97

\bibitem[Weisskopf et al.(1995)]{Weisskopf1995}
Weisskopf, M. C., O'Dell, S. L., Elsner, R. F., \& Van
Speybroeck, L. P. 1995, Proc. SPIE, 2515, 312

\bibitem[Weliachew, Fomalont, \& Greisen(1984)]{Weliachew1984}
Weliachew, L., Formalont, E. B., \& Greisen, E. W. 1984,
\aap, 137, 335

\end{thebibliography}
\end{document}